\documentclass[prc,aps,nofootinbib,showpacs,twocolumn,superscriptaddress]{revtex4}   
\usepackage{epsfig}  
\usepackage{graphicx}  
\usepackage{amssymb}
\usepackage{amsmath}
\usepackage{color}  

\usepackage{dsfont}

\begin{document}  
  
\title{Coupled-channels description of multinucleon transfer and fusion reactions at energies near and far below the Coulomb barrier}  
  
\author{Guillaume Scamps}  
 \email{scamps@nucl.phys.tohoku.ac.jp}  
\affiliation{Department of Physics, Tohoku University, Sendai 980-8578, Japan}

 \author{Kouichi Hagino}  
 \email{hagino@nucl.phys.tohoku.ac.jp}  
 \affiliation{Department of Physics, Tohoku University, Sendai 980-8578, Japan}
\affiliation{Research Center for Electron Photon Science, Tohoku University, 1-2-1 Mikamine, Sendai 982-0826, Japan}
\affiliation{National Astronomical Observatory of Japan, 2-21-1 Osawa, Mitaka, Tokyo 181-8588, Japan}
\begin{abstract}  
We investigate 
heavy-ion multinucleon transfer reactions 
using the coupled-channels formalism. 
We first use the semi-classical approximation and 
show that a direct coupling between the entrance and 
the pair transfer channels improves 
a fit to the experimental one- and two-neutron transfer cross 
sections for the $^{40}$Ca+$^{96}$Zr and $^{60}$Ni+$^{116}$Sn 
systems. 
We then discuss the validity of the perturbative approach 
and highlight the effect of high order terms. 
The effect of absorption is also investigated 
for energies around the Coulomb barrier. 
Finally, we use a quantal coupled-channels approach to achieve 
a simultaneous description of the fusion cross sections and the 
transfer probabilities for the $^{40}$Ca+$^{96}$Zr reaction. 
We find a significant effect of the couplings to the collective 
excited states on the transfer probabilities around the Coulomb 
barrier. 
\end{abstract}

 
\maketitle  

\section {Introduction} 

A transfer of a few nucleons during the reaction process 
plays an important role in heavy-ion 
sub-barrier fusion reactions \cite{Jiang14}. 
The coupling to a transfer degree of freedom 
is in general weaker than 
couplings to collective excitations. However, when the coupling is 
sufficiently large, {\it e.g.,} for a transfer channel 
with a positive $Q$-value, 
it is expected that 
the transfer coupling leads to an extra enhancement of 
fusion cross sections 
below the Coulomb barrier 
\cite{Bro83,Lee84,Esb89,Esb98,Zagrebaev,Tim97,Ste14,Stefanini07,Stefanini13,Bou14,Kohley13,Zhang10,Jia14,Rac14,Sar11, Sar12,Sar13}.

It has been known well that two-neutron transfer reactions 
provide a unique tool to study the pair correlation between 
nucleons \cite{Oer01,VS12,Yoshida62,Pot13}.
Recently two experiments for heavy-ion transfer reactions 
have been carried out at energies far below the Coulomb barrier 
\cite{Cor11,Mon14} in order to study the relation 
between the transfer of one neutron and that of a pair of neutrons. 
See also Refs. \cite{Szilner07,CPS09}. 
The experimental data show an enhancement of 
the two-neutron transfer cross sections compared to 
a simple estimate based on the independent picture, that is, 
a square of the transfer probability of one neutron. 

%the estimated result in the independent particle picture: $P_2=\frac14 P_1^2$ with $P_1$ and $P_2$ the probability to transfer one and two neutrons. 

Several theoretical studies have been performed in order 
to understand the reaction dynamics of the pair transfer process. 
Those include 
the time-dependent perturbation theory based on the 
semi-classical 
approximation \cite{Oer01,VS12,Bro73a,Maglione85,Bro91}, 
the second-order distorted wave Born approximation (DWBA) 
\cite{Pot13,Potel11,Potel13,Bay82}, 
and the semi-classical \cite{CPS09} 
and the quantal \cite{Esb89,Esb98} coupled-channels 
approaches. 
Recently, the time-dependent density functional theory (TDDFT) 
has also been employed 
to investigate the transfer reaction. 
An advantage of this method is that the formalism takes into 
account simultaneously the structure and the dynamics. 
Although 
this method reproduces the individual transfer of 
nucleons \cite{Sim10b,Uma08,Sekizawa13,Eve11}, 
the enhancement factor of a two-neutron transfer probability 
is underestimated with this method \cite{Sca15} 
even if the pairing correlation is taken into account \cite{Sca13}. 
This implies a necessity 
to go beyond the mean-field dynamics with pairing in order 
to take into account all the remaining 
correlations \cite{Sim11,Was09,Ayi15,Wong78}.

In this paper, we carry out a phenomenological study in order to 
understand the transfer dynamics using the formalism of 
the semi-classic and the quantal coupled-channels approaches. 
Our aim in this paper 
is twofold. One is to understand the reaction dynamics 
of the two-neutron transfer reactions. 
We emphasize that the reaction dynamics 
is so complicated that 
many microscopic approaches still lack of prediction for the 
two-neutron transfer. In particular, the nature of the 
two-neutron transfer reaction, 
that is, the relative importance between the 
direct and sequential processes, is still 
under discussions 
\cite{Bay82,Esb98,Pot13,Bro73a,Kam76,Sca13,Par15,Sak01}. 
The second aim of this paper is to describe subbarrier fusion 
reactions using the transfer coupling form factors which are 
consistent with the transfer cross sections. 
For these purposes, 
we phenomenologically adjust the parameters in the form factors 
for the transfer couplings, rather than computing them 
microscopically. 

The paper is organized as follows. In Sec. II, 
we employ the semi-classical time-dependent coupled-channels 
method 
and discuss the 
sequential and the 
direct natures of the two-neutron transfer process. 
In Sec. III, we use the transfer coupling form factors obtained 
in Sec. II to discuss the role of absorption in the two-neutron 
transfer 
reactions. In particular, we investigate the interplay between the 
transfer and the fusion processes using the quantal coupled-channels 
approach. We then summarize the paper in Sec. IV. 

\section{Semi-classical approach}
\label{sec:semi_class}
 
\subsection{Transfer probability}
 
Since the experimental transfer cross sections are often analyzed 
using the semi-classical method \cite{Oer01}, 
we first employ it to investigate 
the nature of the two-neutron transfer. 

With the semi-classical approximation to coupled-channels 
equations, one assumes 
a Rutherford trajectory for 
the relative motion $r(t)$ between the colliding nuclei. 
This yields a time-dependent field for the intrinsic motions in 
the projectile and the target nuclei. 
That is, the nuclear intrinsic wave function is expanded as 
$|\Psi (t)\rangle=\sum_{n=0}^N c_n(t) | n \rangle$, $n=0$ 
corresponding to the entrance channel, and it is then 
evolved in time as 
\begin{align}
i \hbar \frac{d c_n(t)}{dt}=  \sum_{n'}{\cal H}_{nn'}(t)\,c_{n'}(t), 
\label{eq:TDCC}
\end{align}
with $c_n(-\infty)=\delta_{n,0}$. 
Here, ${\cal H}_{nn'}(t)$ is given as 
\begin{equation}
{\cal H}_{nn'}(t)= 
\epsilon_n\delta_{n,n'}+V_{nn'}[r(t)],
\end{equation}
where 
$\epsilon_n$ is the excitation energy for the channel $n$ and 
$V_{nn'}[r(t)]$ is the coupling form factor evaluated along the 
classical trajectory, $r(t)$. 
The probability for the channel $n$ is given as $P_n=|c_n|^2$ 
at $t=+\infty$. 

For the transfer problem, $n$ corresponds to the number of 
transferred nucleons. 
By truncating the transfer channels at $n=2$, 
the coupling Hamiltonian ${\cal H}$ reads,
\begin{align}
   {\cal H}(t) = 
    \left(
    \begin{array}{ccc}
    0 & V_{01}(t) & V_{02}(t) \\
    V_{01}(t) & - Q_1 & V_{12}(t) \\
    V_{02}(t) & V_{12} (t)&  - Q_2
    \end{array}
    \right),
\label{eq:Hamilt}
\end{align}
where $\epsilon_n=-Q_n$ is the 
transfer $Q$-value for each partition, $n$. 
For the coupling form factor 
$V_{nn'}$, we employ 
\begin{align}
 V_{nn'}(r)&=\frac{\beta_{nn'}}{\sqrt{4 \pi} \,a_{nn'}}\, 
              \frac{  \left[ e^{(r-r_p)/a_{nn'}} \right]^3}
{ \left[1+e^{ (r-r_p)/a_{nn'} }  \right]^4 }.
\label{eq:coupling}
\end{align}
This function has the exponential form, 
\begin{align}
V_{nn'}(r)&\sim\frac{\beta_{nn'}}{\sqrt{4 \pi} \,a_{nn'}}\, 
              e^{-(r-r_p)/a_{nn'}},  
\end{align}
for $r\gg r_p$. 
Note that this parametrization differs from a 
frequently used form, $V_{nn'}(r)=\beta_{nn'}\frac{d}{dr} f(r)$, 
where $f(r)$ is a Fermi function, 
only for small values of $r$. 
We find that the solutions for the coupled-channels equations 
are numerically more stable with 
the parametrization given by Eq. (\ref{eq:coupling}). 

In the actual calculations presented below, we use 
$r_p=1.1 \times 
(A_{\rm T}^{1/3} + A_{\rm P}^{1/3} )$ fm, where $A_{\rm T}$ and 
$A_{\rm P}$ are the mass numbers for the target and the projectile 
nuclei, respectively. 
We start the time-dependent Schr\"odinger equation, 
Eq. (\ref{eq:TDCC}), at 
at an initial distance $r_{\rm ini}=30$ fm 
where the coupling is negligible. 
Following Ref. \cite{Bro91}, we evaluate the time dependence of 
the coupling form factor $V_{nn'}(t)$ by averaging the two 
Rutherford trajectories, $r_n(t)$ and $r_{n'}(t)$, as 
$r(t)=(r_n(t)+r_{n'}(t))/2$, where 
$r_n(t)$ and $r_{n'}(t)$ are 
evaluated with 
the energy of $E_n=E_{cm}+Q_{n}$ and $E_{n'}=E_{cm}+Q_{n'}$, respectively. 
In order to avoid a dependence of the results on 
the initial position, $r_{\rm ini}$, 
all the trajectories are arranged so that the distance of the closest 
approach is reached at the same time, $t_{\rm min}$.
We follow the time evolution until the two fragments are 
separated with a distance of 30 fm. We have checked 
that the coupling among the channels are practically negligible 
after this distance, and thus the transfer probabilities are 
not altered. 

In order to reduce 
the number of adjustable parameters, we 
employ two different schemes. 
The first scheme corresponds to a pure sequential process, 
for which we set $V_{02}$ to be zero while $V_{01}$ and 
$V_{12}$ are allowed to be different. 
On the other hand, the second scheme 
corresponds to a direct two-neutron transfer process, for which 
we set $V_{01}(r)=V_{12}(r)$ but $V_{02}\neq 0$ in general. 
The enhancement of two-neutron transfer process is explained 
in a different way with these two schemes. 
In the pure sequential transfer scheme, 
the enhancement factor is due to the increase of 
the probability for the transfer of the second neutron 
after the first neutron is transferred, that is, 
$|V_{12}| > |V_{01}|$. On the other hand, 
in the direct transfer scheme, the enhancement is due 
to the additional coupling between the entrance and the two-neutron 
transfer channels. 

\begin{figure}[tb] 
	\centering\includegraphics[width=\linewidth]{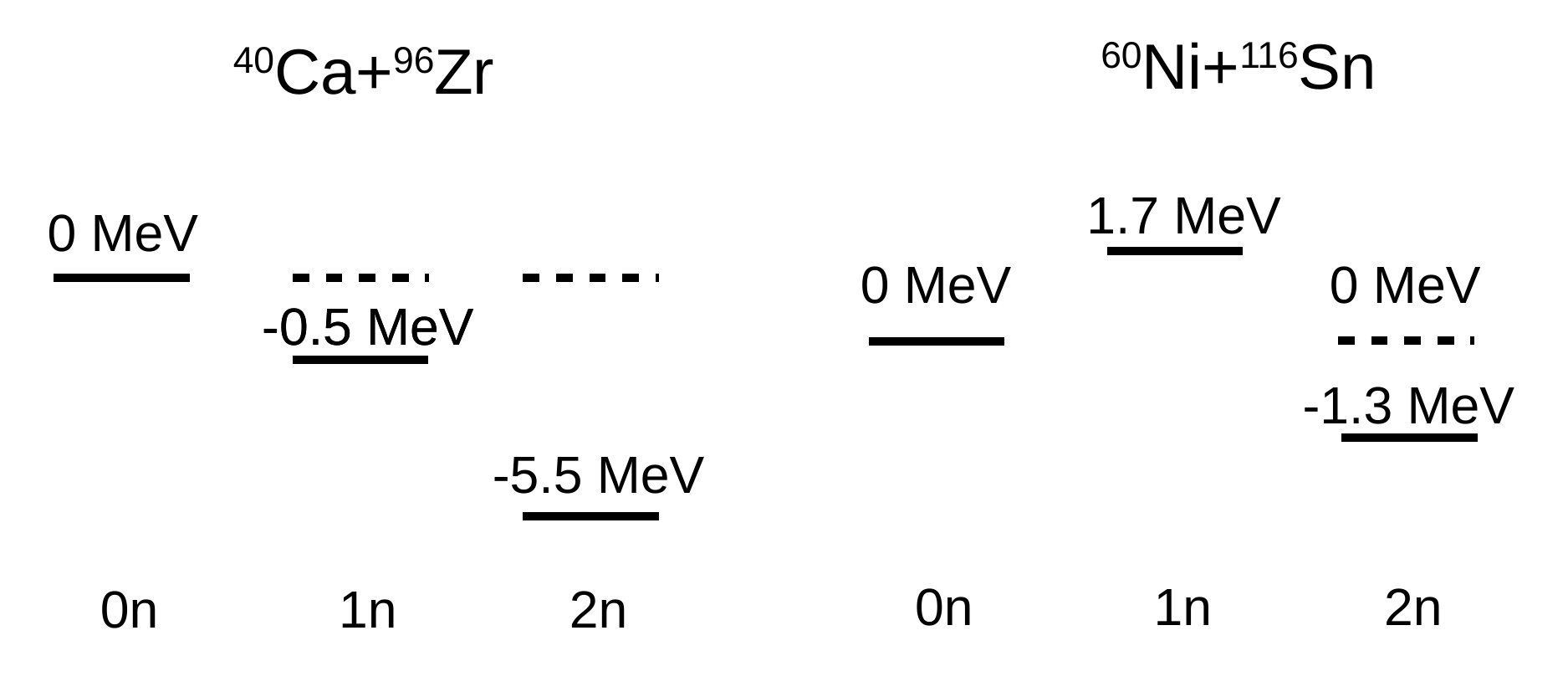}
  	\caption{The 
ground state energies for the one-neutron (1n) and the 
two-neutron (2n) transfer processes for the 
$^{40}$Ca+$^{96}$Zr and $^{60}$Ni+$^{116}$Sn systems. 
The dashed lines denote the optimum $Q$-value for each transfer 
channel.}
\label{fig:states} 
\end{figure}   

We test these schemes on 
the two recent experiments for the 
$^{40}$Ca+$^{96}$Zr and $^{60}$Ni+$^{116}$Sn systems 
\cite{Cor11,Mon14}. In these references, the experimental data, 
taken at backward angles in the center of mass frame, are given 
in terms of the transfer probability defined by the ratio of the 
transfer cross sections to the Rutherford cross sections, 
$P_{xn}\equiv d\sigma_{xn}/d\sigma_R$, with $x=1,2,\cdots$, 
as a function of the distance 
of the closest approach, $D$, for the Rutherford trajectory. 
Figure \ref{fig:states} summarizes the ground state energies 
for these systems. The $Q$-value for the ground state to 
the ground state transition, $Q_{\rm gg}$, is positive both for the 
one-neutron and the two-neutron transfer channels for the 
$^{40}$Ca+$^{96}$Zr system, while it is 
negative for the one-neutron transfer and positive for the two-neutron 
transfer for the 
$^{60}$Ni+$^{116}$Sn system. 

\begin{figure}[tb] 
\centering\includegraphics[width=\linewidth]{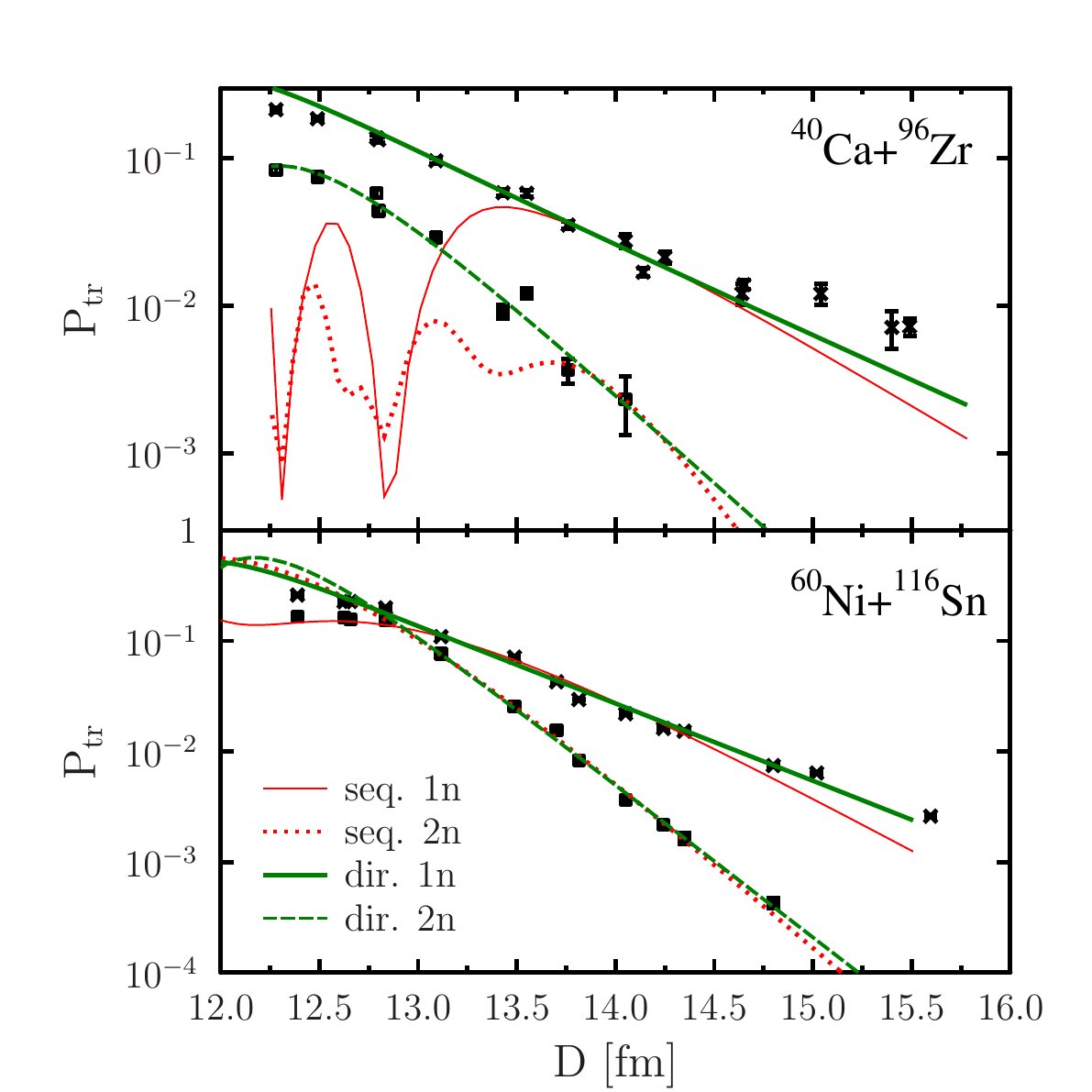}
\caption{(Color online) The transfer probabilities 
for the $^{40}$Ca+$^{96}$Zr (the upper panel) 
and $^{60}$Ni+$^{116}$Sn (the lower panel) reactions 
as a function of the distance of the closest approach, $D$, 
for the Rutherford trajectory. 
The experimental data, taken form Refs. \cite{Cor11,Mon14}, for 
the one-neutron (the crosses) and the two-neutron (the squares) 
transfers 
are compared to transfer probabilities for the ground state to 
the ground state transitions obtained with the 
pure sequential scheme (the thin solid line for the 
1n transfer and the dotted line for the 2n transfer) and 
with the direct transfer scheme (the thick solid line for 
the 1n transfer and the dashed line for the 2n transfer). 
} 
\label{fig:res_Qgs_gs} 
\end{figure}   

\begin{table}
\caption{The parameters for the transfer coupling form factors 
for the $^{40}$Ca+$^{96}$Zr reaction. For all the rows, 
$a_1$ and $\beta_1$ are those for the coupling between the entrance 
and the 1n transfer channels. 
For the sequential scheme (seq.), 
$a_2$ and $\beta_2$ are those for the coupling between 
the 1n and the 2n transfer channels, that is, 
$a_2=a_{12}$ and $\beta_2=\beta_{12}$. 
For the direct transfer scheme (dir.), 
$a_2$ and $\beta_2$ are those for the coupling between 
the 0n and the 2n transfer channels, that is, 
$a_2=a_{02}$ and $\beta_2=\beta_{02}$. 
These parameters are obtained either with the ground state to the 
ground state transfer $Q$-values ($Q_{\rm gg}$) or with the 
optimum $Q$-values ($Q_{\rm opt}$). 
}
\begin{tabular}{c|c|c|c|c|c}
\hline\hline
$Q$ & Scheme  & $a_1$ (fm) &  $\beta_1$ (MeV\,fm) &   
$a_2$ (fm) & $\beta_2$ (MeV\,fm)   \\
 \hline
$Q_{\rm gg}$ & seq. & 1.015 & 121.8 & 0.726 & 4470     \\
$Q_{\rm gg}$ & dir. & 1.309 & 39.05 & 0.727 & 443 \\
$Q_{\rm opt}$ & seq. & 1.113 & 80.33  & 1.458 & 107.4  \\
$Q_{\rm opt}$ & dir. & 1.230 & 51.56 & 0.700 & 278.9 \\
\hline\hline
\end{tabular}
\label{tab:coupl_CaZr_param}
\end{table}

\begin{table}
\caption{ Same as Table \ref{tab:coupl_CaZr_param}, but 
for the $^{60}$Ni+$^{116}$Sn reaction.}
\begin{tabular}{c|c|c|c|c|c}
\hline\hline
$Q$ & Scheme  & $a_1$ (fm) &  $\beta_1$ (MeV\,fm) &   
$a_2$ (fm) & $\beta_2$ (MeV\,fm)   \\
 \hline
$Q_{\rm gg}$ & seq. & 0.877  & 138.9  & 1.36  & 131.9     \\
$Q_{\rm gg}$ & dir. & 1.18 & 40.6 & 0.602 & 363.4 \\
$Q_{\rm opt}$ & seq. & 0.893  & 126.6 & 1.38  & 109.4   \\
$Q_{\rm opt}$ & dir. &  1.13 & 47.1 & 0.600 & 384.2 \\
\hline\hline
\end{tabular}
\label{tab:coupl_NiSn_param}
\end{table}

Let us first consider a case with the ground state to the ground 
state $Q$-value, that is, a case with $Q_n=Q_{\rm gg}(n)$ in Eq. 
(\ref{eq:Hamilt}). 
We adjust the coefficients $a_{nn'}$ and $\beta_{nn'}$ 
in Eq. (\ref{eq:coupling}) by fitting the experimental data 
for $D>13.5$ fm, for which the coupling is weak and 
only the first order dynamics is important. 
Figure \ref{fig:res_Qgs_gs} shows a comparison of the calculated 
transfer probabilities so obtained with 
the experimental data. 
Those are obtained by varying the 
center of mass energy with fixed values of the scattering angle, 
that is, $\theta_{\rm c.m.}=$ 140 degrees for the 
$^{40}$Ca+$^{96}$Zr and $^{60}$Ni+$^{116}$Sn systems. 
For each energy, the impact parameter 
is determined so that the scattering angle for the Rutherford 
trajectory for the entrance channel 
is consistent with the given value of the scattering angle. 
In the figure, the thin solid and the dotted lines 
denote the results with the 
pure sequential scheme for the 1n and 2n transfer processes, 
respectively. 
On the other hand, the thick solid and the dashed lines 
denote those with the direct transfer scheme. 

The values of the parameters are summarized in Tables I and II 
for the 
$^{40}$Ca+$^{96}$Zr and $^{60}$Ni+$^{116}$Sn systems, respectively. 
Notice that the direct two-neutron 
transfer has a shorter range with 
small $a_2$ than the sequential coupling, in accordance with the 
argument in Ref.  \cite{Esb14}.
One can also notice that 
with the sequential scheme one has to take a considerably different 
parameters for $V_{12}$ from those for $V_{01}$ 
in order to reproduce the experimental two-neutron transfer 
probability. 
This may appear unnatural since a drastic change of structure 
is not expected even after one neutron is transferred, 
although it may simply mock up a constructive interference 
of different transfer paths \cite{LFV14}. 

In Fig. \ref{fig:res_Qgs_gs}, an unexpected behavior can be seen 
for the $^{40}$Ca+$^{96}$Zr reaction with the pure sequential 
scheme (the dotted and the thin solid lines). 
That is, the transfer probabilities oscillate as a function 
of $D$, even though the exponential tail is correctly reproduced. 
This behavior is due to 
the large $Q$-value for the two-neutron 
transfer channel, $Q_{2}$=5.5 MeV. 

\begin{figure}[tb] 
\centering\includegraphics[width=\linewidth]{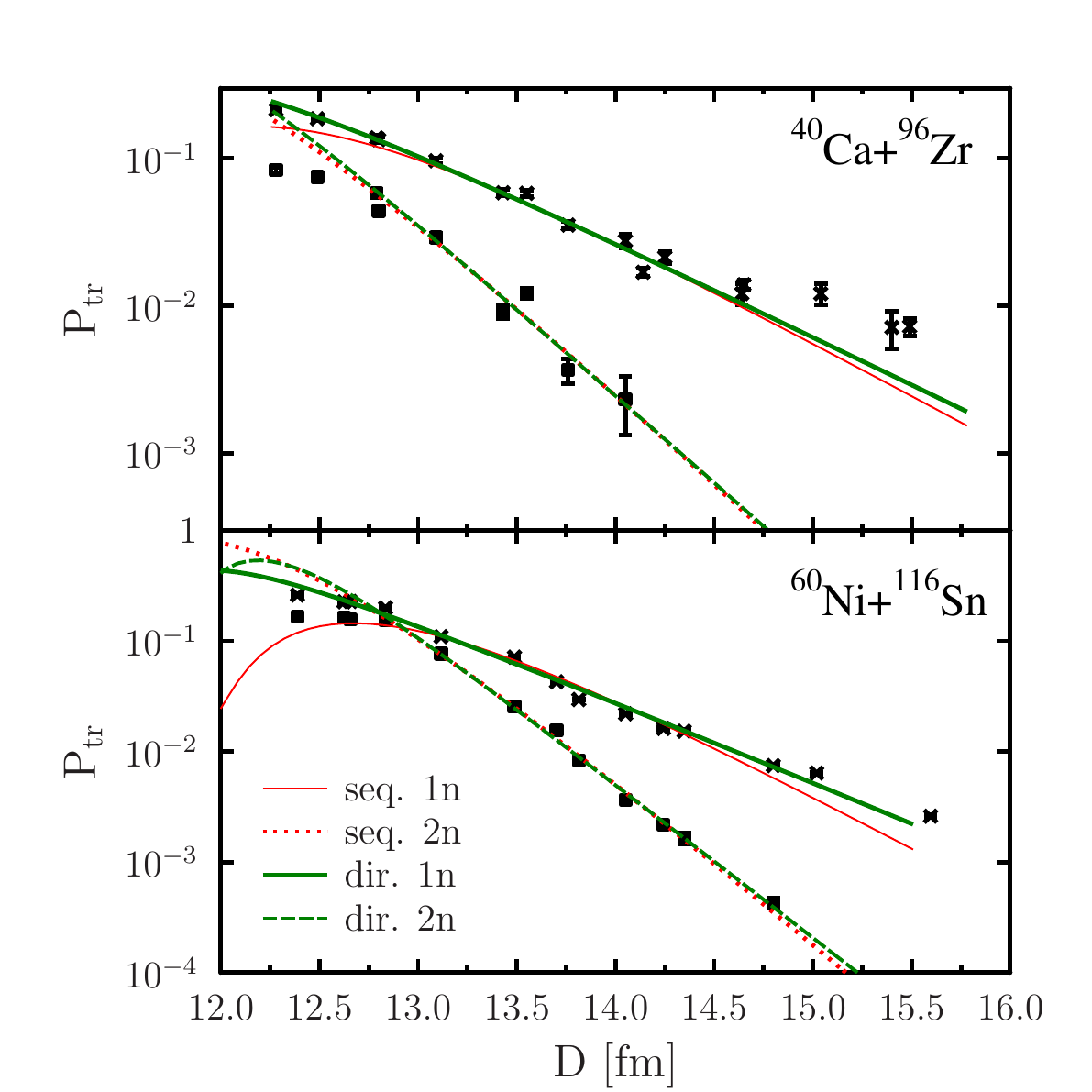}
\caption{(Color online)  Same as Fig. \ref{fig:res_Qgs_gs}, but 
with the optimum $Q$-values (see the text).} 
\label{fig:res_Q0} 
\end{figure}   

In reality, however, the transfer takes place mainly 
to excited states, rather than to the 
ground state \cite{Cor11,Rowley01}. In fact, the optimum $Q$-value 
for a neutron transfer is $Q_{\rm opt}=0$ \cite{Henning78}, 
and the coupling to the ground state is 
much weaker \cite{Zagrebaev}. 
Figure \ref{fig:res_Q0} is obtained by using the 
optimum $Q$-values, that is, $Q$=min(0,$Q_{\rm gg}$) 
(see Fig. \ref{fig:states}). 
See Tables I and II for the parameters.  
As expected, the agreement with the experimental data is much 
improved. 

In Fig. \ref{fig:res_Q0}, 
one can notice that 
the one-neutron transfer probability is well reproduced 
with the direct transfer scheme. In contrast, a significant deviation 
is seen for small values of $D$ with the sequential scheme. 
This can be attributed to the sequential nature of the 
transfer dynamics. 
For large values of $D$, the transfer probabilities are small 
and the sequential transfer to the 2n channel 
does not modify $P_1$. However, as $D$ decreases, 
the two-step process becomes significant, reducing 
the probability for the one-neutron transfer channel. 
With the direct two-neutron transfer scheme, on the other hand, 
the two-neutron transfer takes place mainly from the entrance 
channel (see Sec. II-C below), as the coupling between the 1n and 
the 2n channel is kept to be relatively weak. 

\begin{figure}[tb] 
\centering\includegraphics[width=\linewidth]{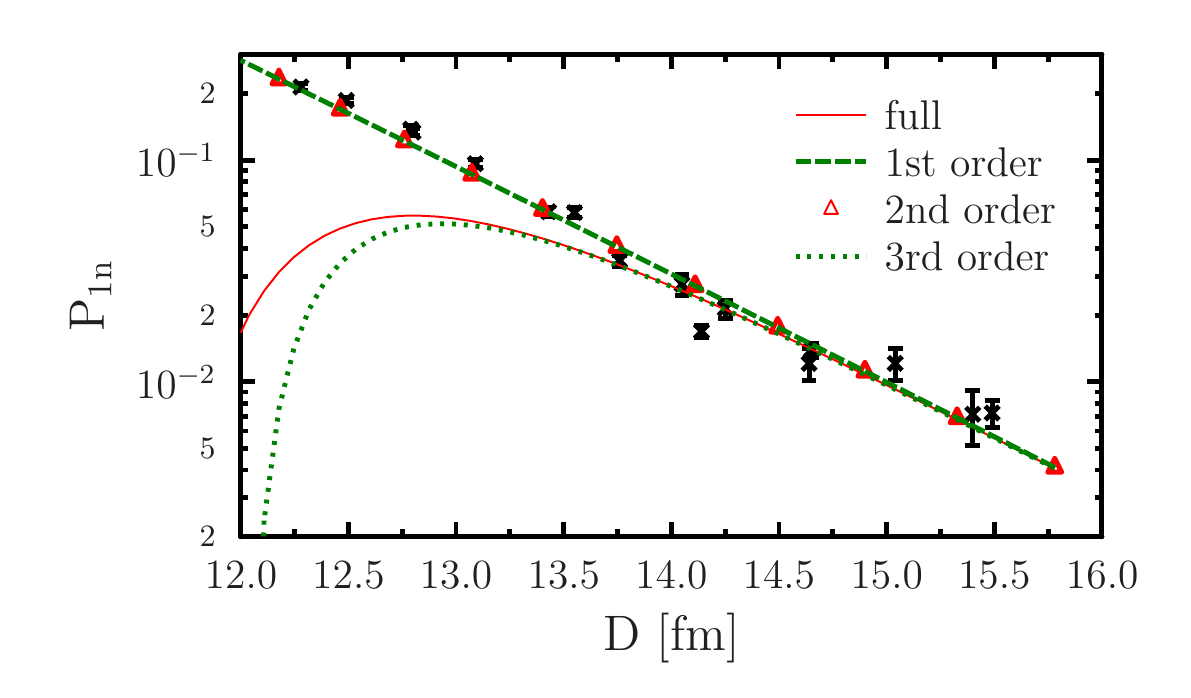}
\caption{(Color online) The one-neutron transfer probability 
as a function of the distance of the closest approach 
for the $^{40}$Ca+$^{96}$Zr reaction. 
The sequential two-neutron scheme with the optimum $Q$-value is 
employed. 
The dashed line, the triangles, and the dotted line are the results 
of the time-dependent perturbation theory with the first, the second, 
and the third orders, respectively. 
Those results are compared to the exact solution (the solid line) 
and to the experimental data (the crosses). 
The coupling parameters used here are  $a_1=$1.55 fm, 
$\beta_1=$ 26.4 MeV\,fm, $a_2=$1.286 fm and $\beta_2=$150.4 MeV\,fm. 
}
\label{fig:TDPT} 
\end{figure}   
 
In order to better understand this phenomena, one can use the 
time-dependent perturbation theory by separating the contribution of 
each order. 
Figure \ref{fig:TDPT} shows 
the approximate solutions 
for the one-neutron transfer reaction 
for the $^{40}$Ca+$^{96}$Zr system obtained with the 
sequential two-neutron scheme. Here, the coupling parameters are 
readjusted in order to reproduce the experimental data for large $D$ 
with the first order dynamics.
Up to the first order, 
the one-neutron transfer probability 
is found to be exponential with the distance of the 
closest approach 
in the whole range of $D$ 
shown in the figure. As a consequence, 
the calculation appears to be consistent with the experimental 
data. This remains to be the same up 
to the second order, since there is no 
second order contribution to the one-neutron transfer channel. 
On the other hand, if one considers up to the third order, the 
result is drastically changed and becomes close to the full order 
result obtained by solving the time-dependent coupled-channels 
equations, Eq. (\ref{eq:TDCC}), without using the perturbation 
theory. 
That is, the third order process now includes 
the diminution of the one-neutron transfer probability due 
to the second order process to the 2$n$ channel. 
From this comparison, we can conclude that the perturbative 
calculations must be used with caution for an 
application to multinucleon transfer calculations, 
since a good agreement with experimental data 
may be an artifact of the first order
perturbation theory. 

Notice that the sequential scheme largely underestimates the transfer 
probabilities at small $D$. 
In this region, the absorption effect would be important, but the effect of 
absorption will always decrease the transfer probabilities, 
as we will discuss in Sec. II-C. Evidently, 
a pure sequential two-neutron transfer is not compatible 
with the experimental data, at least with this simple model. See also 
Refs. \cite{Par15,Esb98} for a similar 
conclusion. 
We have confirmed that this conclusion remain the same 
even if we include a few states around the optimum $Q$-value for 
each transfer partition. 

Notice that 
the semi-classical transfer calculations have indicated that 
the two-neutron transfer process occurs predominantly 
with a sequential process \cite{Bro73a,Maglione85,Bro91}.  
The present result is somewhat in contradiction with 
the previous finding of the semi-classical method. 
The difference may be due to the fact that the previous semi-classical calculations 
take into account many intermediate 1n transfer channels 
with a wide distribution of excitation energy 
while we include only a single (or at most 
a few) channel. 
In this sense, our coupling scheme may have to be regarded as an effective one, which 
implicitly takes into account the effect of many other intermediate channels. 
It would be an interesting future work to clarify how the elimination of the 
intermediate states gradually changes the nature of the two-neutron transfer 
couplings. 

%The difference between the present contribution and the precedent work is that here no microscopic assumption is made.

%\subsection{Validity of the perturbative treatment}

\subsection{Nature of two-neutron transfer}
 
\begin{figure}[tb] 
\centering\includegraphics[width=\linewidth]{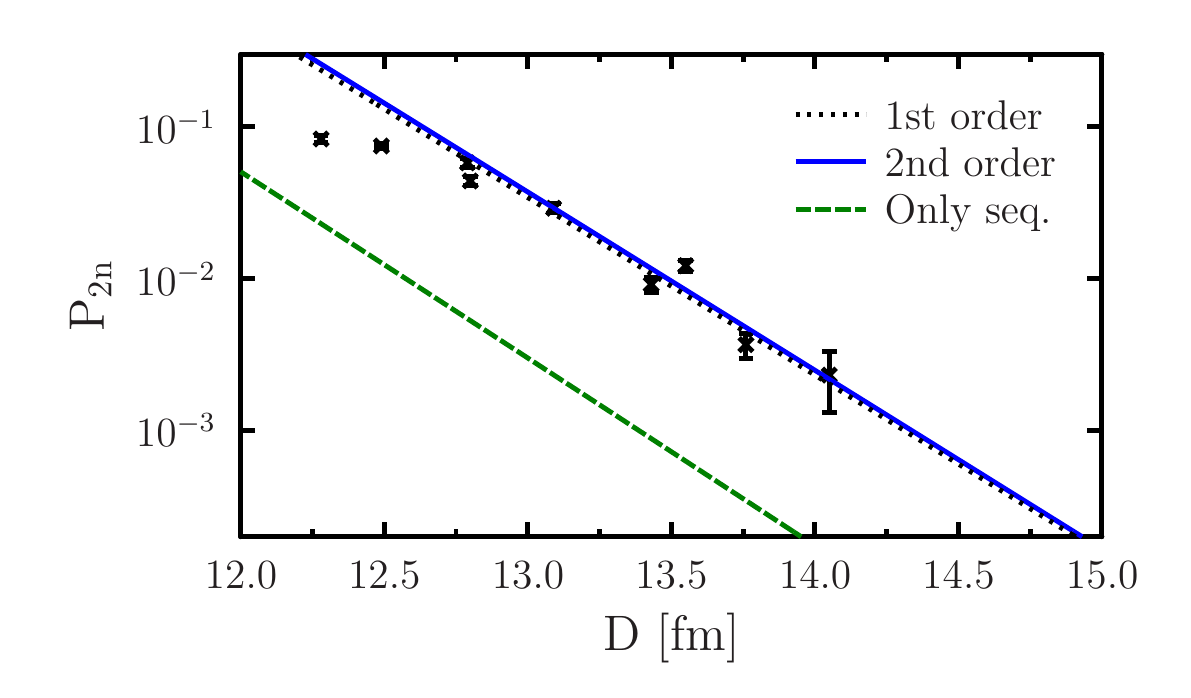}
\caption{(Color online) The two-neutron transfer probability 
for the $^{40}$Ca+$^{96}$Zr system 
as a function of the distance of the closest approach, $D$. 
The direct two-neutron transfer scheme with the optimum $Q$-values is employed. 
The results of the time-dependent perturbation theory up to the first and the second orders 
are shown by the dotted and the solid lines, respectively. 
The result of the second order calculation 
neglecting the direct two-neutron transfer couplings 
is also shown by the 
dashed line (see text). The experimental data are taken from Ref. \cite{Cor11}.}
\label{fig:sim_vs_seq} 
\end{figure}   
 
In the previous subsection, we have argued that the direct two-neutron transfer 
scheme is more plausible 
than the pure sequential transfer scheme as long as a simple coupling scheme as in 
Eq. (\ref{eq:Hamilt}) is employed. Let us then investigate 
the nature of the two-neutron transfer process by assuming the direct transfer 
scheme with the optimum $Q$-values. 
Figure \ref{fig:sim_vs_seq} shows the result for the probability of the two-neutron 
channel for the $^{40}$Ca+$^{96}$Zr system obtained by the perturbation theory. 
To this end, we use the same parameters as those used in 
Fig. \ref{fig:res_Q0}. 
The dotted line denotes the result of the first order perturbation theory, which 
includes only the direct population of the 2n transfer channel from the entrance channel. 
The solid line, on the other hand, denotes the result of the second order 
perturbation theory, which 
in addition 
includes the sequential two-neutron transfer via the intermediate 1n transfer channel. 
For a comparison, the figure also shows the calculation with the second order 
contribution only (the dashed 
line), that is, the calculation neglecting the direct two-neutron coupling. 
One can see that the sequential process is negligibly small as compared to the first order 
process. That is, the two-neutron transfer in this model 
is largely dominated by the direct two-neutron transfer 
from the entrance channel partition. 
Notice that this conclusion is a direct consequence of the simplified 
transfer model with the choice of $V_{12}(r)=V_{01}(r)$, and it would need to be confirmed with a more general model.

\subsection{Effect of absorption}

While the experimental data for the two-neutron transfer are well 
described with the direct two-neutron 
transfer 
scheme for the distance of the closest approach of 13 fm or larger, 
a discrepancy is found for  $D<13$ fm 
(except for the $^{40}$Ca+$^{96}$Zr system with $Q_{\rm gg}$). 
While the calculations predict a nearly exponential behavior of $D$, 
the experimental data show a much slower dependence in this region. 
This could be attributed to 
several mechanisms beyond the present model, such as a deviation 
of the classical trajectory 
from the Rutherford trajectory, a loss of one-to-one correspondence 
between the experimental and the 
theoretical definitions of the transfer probabilities (that is, 
the quantity $d\sigma_{xn}/d\sigma_R$ may not 
be able to be interpreted as a transfer probability at small values of $D$), and a 
deviation of the coupling form factor from that 
given by Eq. \eqref{eq:coupling}. 

 \begin{table}
\caption{The parameters of the absorbing potential for each channel 
required to fit the experimental data. }
\begin{center}
\begin{tabular}{c|c|ccc}
\hline\hline
System & Channel & $W_0$ (MeV) & $R_W$ (fm) & $a_W$ (fm) \\
\hline
 $^{40}$Ca+$^{96}$Zr & 0n and 1n & 9.96 & 11.88 & 0.12 \\
& 2n & 9.89 & 12.10 & 0.18 \\
\hline
$^{60}$Ni+$^{116}$Sn & 0n and 1n & 10.03 & 12.01 & 0.08 \\
& 2n & 10.08 & 12.25 & 0.16 \\
\hline\hline
\end{tabular}
\end{center}
\label{tab:param_fus1}
\end{table}
 
\begin{figure}[tb] 
\centering\includegraphics[width=\linewidth]{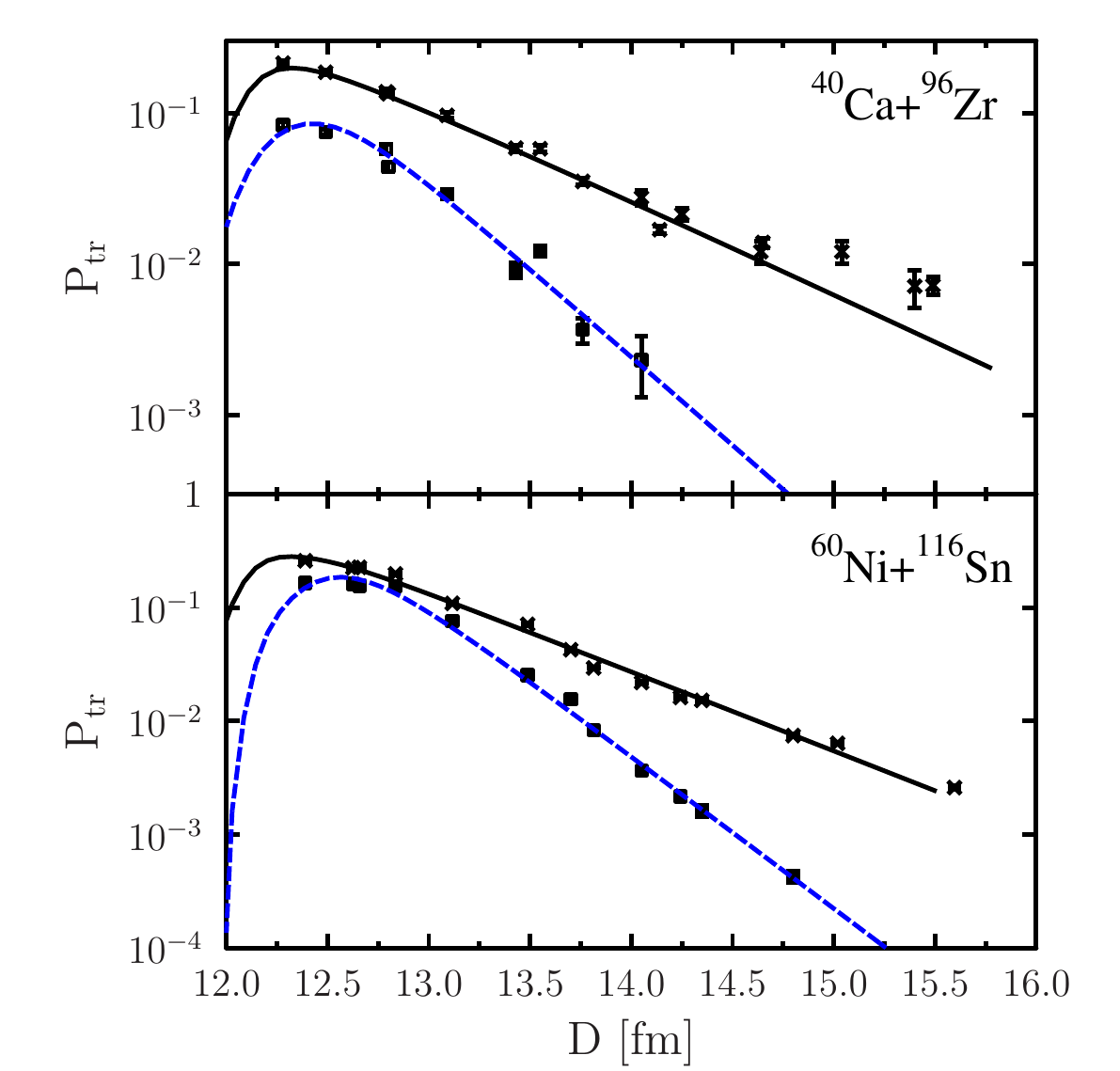}
\caption{(Color online) The transfer probabilities as a function of the 
distance of the closest approach, $D$, calculated by including the effect of 
absorption of the classical trajectories with the imaginary potential. 
The direct two-neutron transfer scheme with the optimum $Q$-value 
is employed. The solid lines denote the results for the one-neutron transfer process, 
while the dashed lines are for the two-neutron transfer process. 
The experimental data are taken from Refs. 
\cite{Cor11,Mon14}.}
\label{fig:sim_Q0_fus} 
\end{figure} 
 
In the following, we choose to test a hypothesis that this 
effect originates from the absorption of 
the wave function, which corresponds to a capture and/or inelastic excitations. 
In order to take into account this effect in a simple way, we add an imaginary potential to 
the diagonal part of the Hamiltonian, Eq. \eqref{eq:Hamilt}, with the expression,
\begin{align} 
iW(r)=\frac{-iW_0}{1+\exp[(r-R_W)/a_W]}, 
\label{eq:abs_pot}
\end{align}
for which the parameters $W_0$, $R_W$, and $a_W$ may be different for each channel. 
We adjust these parameters in order to reproduce the experimental transfer 
probabilities using the direct two-neutron transfer scheme with the optimum $Q$-values. 
The resultant values for the parameters 
are summarized in Table III, and the fit to the experimental 
transfer probabilities is shown in Fig. \ref{fig:sim_Q0_fus}.
One can see that a good agreement between the calculations and the experimental 
data is achieved, including the region with small values of the distance of the closest 
approach, although 
the good reproduction of the data may be due to 
the large number of adjustable parameters. 

\begin{figure}[tb] 
\centering\includegraphics[width=\linewidth]{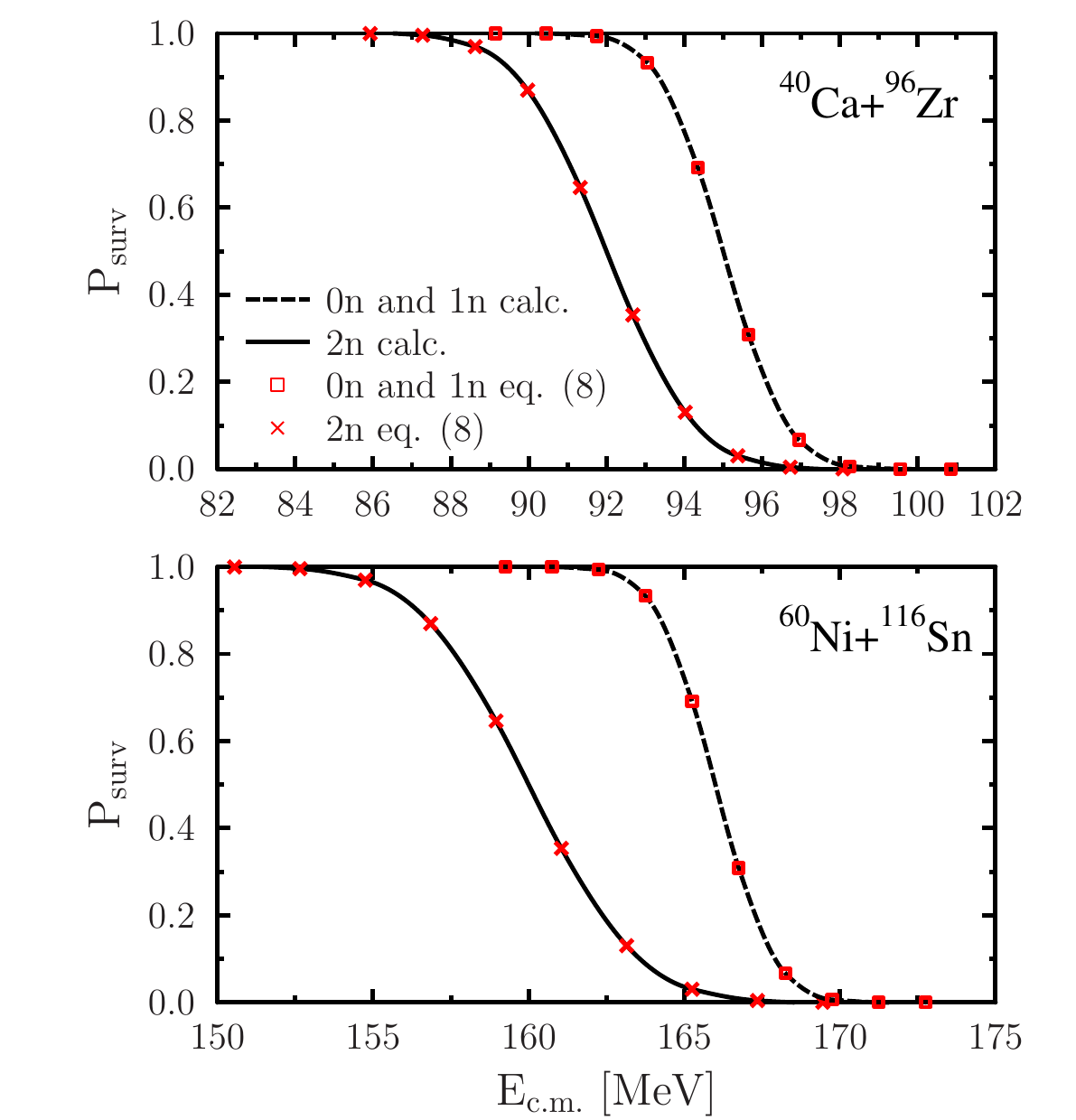}
\caption{(Color online) %\error FIGURE HAS BEEN CHANGED
 The $s$-wave survival probability for each channel 
as a function of the center of mass energy. It is obtained by solving the time-dependent Schr\"odinger equation 
with a single channel including the imaginary potential. 
The crosses and the squares show a fit with the 
the complementary error function given by Eq. \eqref{eq:P_surv}. }
\label{fig:sim_Q0_fus_prob} 
\end{figure} 
 
\begin{table}[!ht] 
\caption{The parameters for the 
complementary error function given by Eq. \eqref{eq:P_surv} which fit the results of 
the time-dependent calculations for the survival probability for each channel. } 
%\begin{center}
\begin{tabular}{c|c|c|c}
\hline\hline
System & Channel  & $B$ (MeV)& $\sigma$ (MeV) \\
 \hline
  $^{40}$Ca+$^{96}$Zr & 0n  &   95   &  1.3  \\
 & 1n  & 95   &  1.3 \\
 & 2n  &   92  & 1.8 \\
 \hline
 $^{60}$Ni+$^{116}$Sn& 0n & 166 & 1.3 \\
 & 1n  & 166 & 1.5  \\
 & 2n  & 160 & 2.8 \\
\hline\hline
\end{tabular}
%\end{center}
\label{tab:param_fus2}
\end{table}
 
The single-channel calculations with those absorbing potentials 
lead to the survival probability for each channel shown in 
Fig. \ref{fig:sim_Q0_fus_prob}. 
Notice that this is not the survival probability for the whole transfer 
process, that is, $P_{\rm surv} = |c_0|^2+|c_1|^2+|c_2|^2$, 
but the single-channel probability corresponding to that 
given by \cite{Bro91}  
\begin{equation}
P^{(xn)}_{\rm surv}(E)=1-\exp\left(\frac{2}{\hbar}\int^\infty_0W[r_n(t)]dt\right). 
\end{equation}
We find that the survival probability for each channel, $P^{(xn)}_{\rm surv}$, can be well 
parametrized with the complementary error function, 
\begin{align} 
P^{(xn)}_{\rm surv}(E) \sim  
\frac12 +  \frac12  {\rm erfc} \left( \frac {E-B}{\sqrt2 \sigma} \right), 
\label{eq:P_surv}
\end{align}
with the parameters given in Table IV (see Fig. \ref{fig:sim_Q0_fus_prob}). 
Notice that the values of the parameters are considerably different between the 
two-neutron transfer channel and the one-neutron transfer channel. 
The meaning of the imaginary potential is to take into account the loss of flux from 
the model space explicitly taken into account in the calculations 
to the outside, which include 
the tunneling through the barrier (that is, the capture) and the inelastic processes. 
In either case, the imaginary potential reduces the transfer probabilities. 
The different values of the parameters for each channel implies that the effect of 
absorption may be different in each channel. 
We find that it is essential to have the channel dependence in the parameters of the 
absorbing potential in order to account for 
the diminution of the 2n transfer probability without significantly 
degrading the reproduction of the 1n transfer channel. 

%The meaning of this imaginary potential is to take into account the tunneling through the barrier as well as the inelastic process of the colliding nuclei that is not taken into account due to the classical trajectory of the two fragments. This imaginary potential absorbs a part of the wave function, the transfer probabilities are then reduced. At the first order, we can write, 
%%
%\begin{align} 
%P_n=P_{\rm surv} P'_n,
%\end{align} 
%%
%with $P'_n$ the transfer probabilities without absorbing potential. 
%Then, if the absorption is the same in each transfer channel, it is not possible to reproduce the diminution of the 2n probability without degrading the reproduction of the single transfer channel. It is then necessary to have different parameters $a_i$ for each channel corresponding to different barrier position. 

%The difference of barrier height (see Tab. \ref{tab:param_fus2}) for the 1n and 2n transfer channels can be interpreted as a complex dynamical mechanism. Nevertheless, to understand this mechanism a reaction model that treats simultaneously the transfer and the fusion mechanism should be used.
 
\section{Quantal Coupled-channels approach}
\label{sec:CC}
 
We have shown 
in the previous section 
that the role of absorption may differs among the different transfer channels.  
We have achieved this conclusion using the semi-classical method. 
A drawback of the semi-classical method is that the tunneling process is not 
easy to be described with it, although one may still do it using the time-evolution 
along the imaginary time axis \cite{HT12,THM95}. 
For this reason, in this section, we shall use the full quantal coupled-channels 
approach, which can also be applied to a simultaneous description of fusion cross 
sections and transfer probabilities. 

The coupled-channels equations for a total angular momentum $J$ read \cite{HT12},
\begin{widetext}
\begin{align}
\left[- \frac{\hbar^2}{2 \mu} \frac{d^2}{dr^2} + \frac{J(J+1)\hbar^2}{2 \mu r^2}  
+ V_N(r) + i W_N(r) + \frac{Z_P Z_T e^2}{r}  + \epsilon_n - E \right] u_n(r) + 
\sum_m V_{nm}(r) u_{m}(r) = 0, 
\end{align}
\end{widetext}
where $\mu$ is the reduce mass, $E$ is the center of mass energy, 
and $Z_P$ and $Z_T$ are the charge numbers of the projectile 
and the target, respectively.  
We have used the iso-centrifugal approximation \cite{HT12} and assumed that the angular 
momentum does not change for each channel. 
We use a Woods-Saxon parametrization for the nuclear potential, $V_N$, that is, 
\begin{align}
V_N(r)=\frac{-V_0}{1+\exp[(r-R_0)/a_0]}
\end{align}
with  $R_0=r_0 (A_P^{1/3}+A_T^{1/3})$. 
We use the Woods-Saxon parametrization also for the imaginary potential, $W_N$, 
as in Eq. \eqref{eq:abs_pot}. 

We solve the coupled-channels equations using a modified version of the 
computer program {\tt CCFULL} \cite{Hag99} in order to construct the transfer 
cross sections from the $S$-matrix. The transfer probability is then calculated 
using the same definition as the experimental probability, that is, 
$P_{xn}\equiv d\sigma_{xn}/d\sigma_R$. 
 
\subsection{Quantum effect }
 
\begin{figure}[!ht] 
	\centering\includegraphics[width=\linewidth]{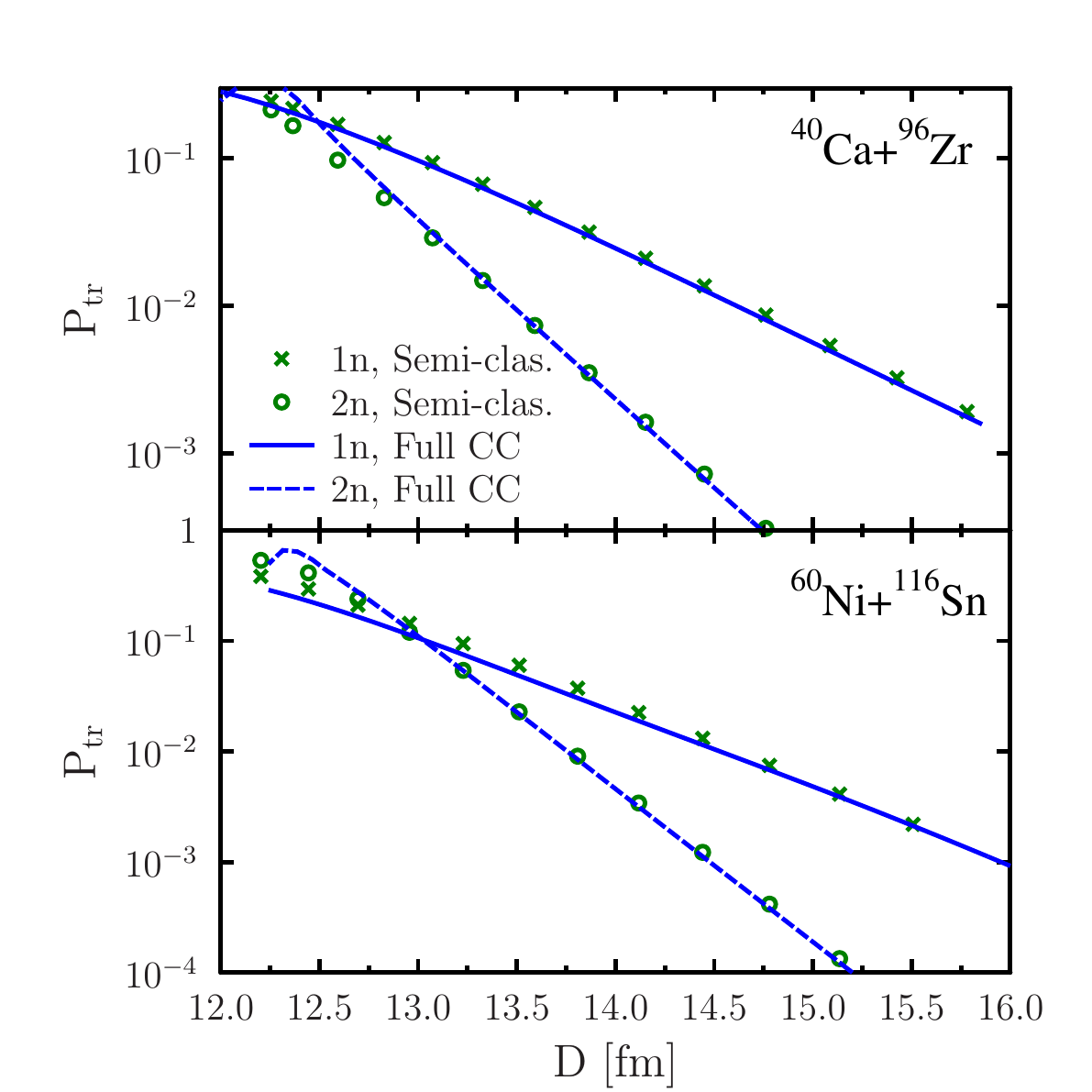}
  	\caption{(Color online) A comparison between the 
transfer probabilities 
obtained with the semi-classical approximation (the crosses and the 
circles) and those with the quantal coupled-channels calculations 
(the solid and the dashed lines) for the $^{40}$Ca+$^{96}$Zr 
and $^{60}$Ni+$^{116}$Sn systems. }
\label{fig:comp_SC_cc_ratio} 
\end{figure} 
 
\begin{table}
\caption{The parameters for the nuclear potential used in the coupled-channels 
calculations for each system. } 
\begin{center}
\begin{tabular}{c|c|c|c|c|c|c}
\hline\hline
System &$V_0$ & $r_0$ & $a_0$ & $W_0$ & $r_W$ & $a_W$ \\
& (MeV) & (fm) & (fm)& (MeV) & (fm) & (fm) \\
 \hline
$^{40}$Ca+$^{96}$Zr & 87.0 & 1.13 & 0.700 & 40 & 1.1 & 0.2   \\
$^{60}$Ni+$^{116}$Sn &  80.0 & 1.1 & 0.487 & 20 & 0.9 & 0.487\\
\hline\hline
\end{tabular}
\end{center}
\label{tab:comp_SC_cc_ratio}
\end{table}

Let us first solve the coupled-channels equations including only the transfer channels,  
and examine the validity of the semi-classical trajectory approximation. 
To this end, 
we use the direct two-neutron transfer scheme with the optimum $Q$-values, 
for which the parameters for the transfer coupling form factors are given in 
Tables \ref{tab:coupl_CaZr_param} and \ref{tab:coupl_NiSn_param}. 
We use the parameters in Table \ref{tab:comp_SC_cc_ratio} for 
the nucleus-nucleus potential (both for the real and the imaginary parts).
 In order to simplify the comparison between the 
semi-classical and the quantal calculations, 
we neglect the effect of absorption on the transfer probabilities. 
For this purpose, the real and the imaginary parts of the nuclear 
potential are not taken into account in the semi-classical calculation. 
Moreover, we choose the parameters of the nuclear potential 
so that they yield a higher barrier than the systematics. 

Figure \ref{fig:comp_SC_cc_ratio} shows a comparison between the 
quantal coupled-channels calculations and the semi-classical coupled-channels 
calculations for the same coupling potentials. One can find a good agreement 
between the two calculations for the distance of the closest approach of 
$D>13.5$ fm, that is, the two calculations differ only by about 10 \%.  
%which is close to the numerical error of the calculations. 
The difference between the two calculations is more significant for  $D<13$ fm, 
where the quantum effect as well as the absorption process play an important 
role. 
 
Evidently, the conclusions obtained in the previous section 
with the semi-classical method remain unchanged even if 
we use the quantal coupled-channels method. 

\subsection{Role of absorption}

\begin{figure}[!ht] 
\centering\includegraphics[width=\linewidth]{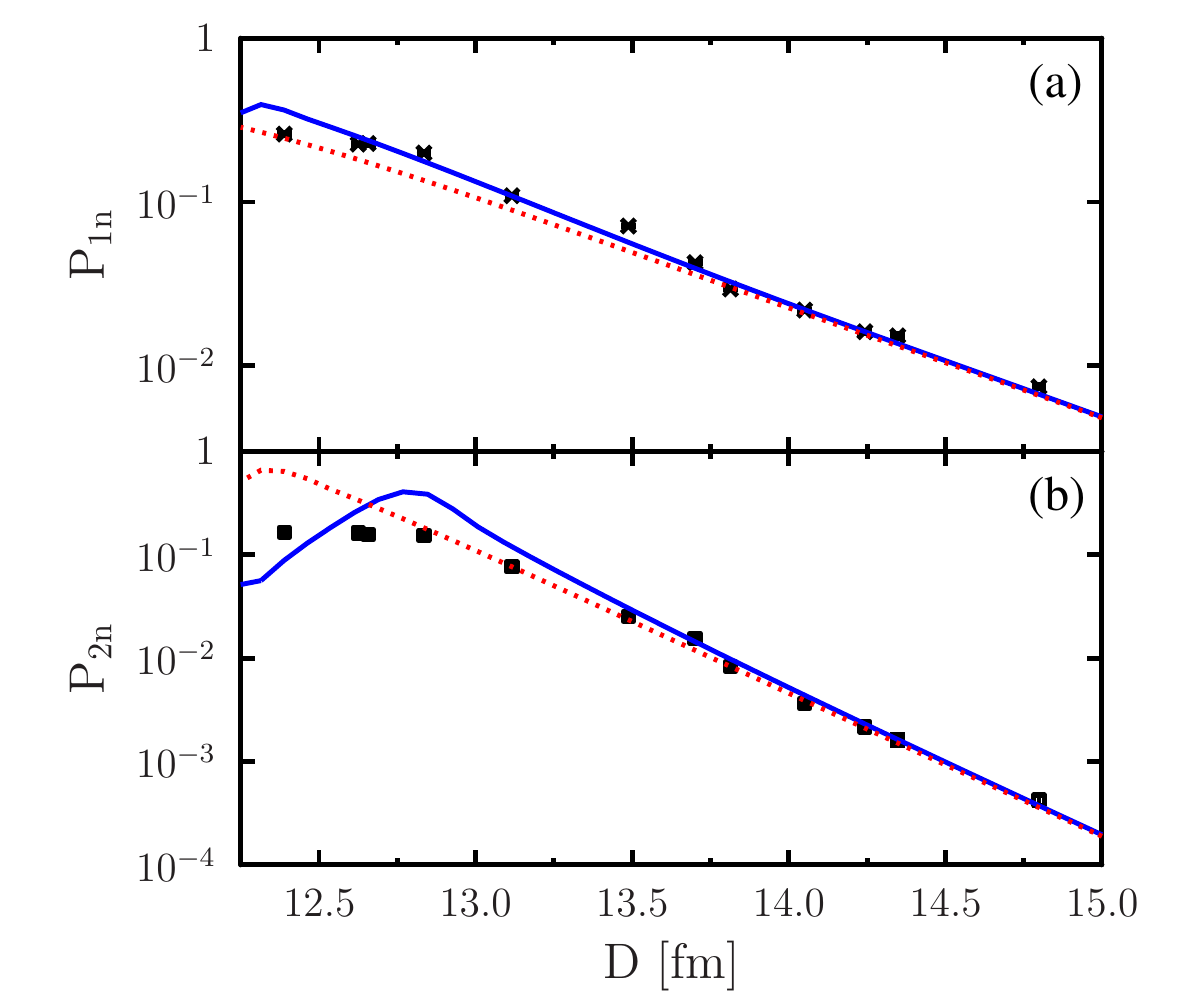}
\caption{(Color online)  The one-neutron transfer (the upper panel) and the two-neutron transfer (the lower 
panel) probabilities as a function of the distance of the closest approach for the $^{60}$Ni+$^{116}$Sn reaction. 
The solid and the dotted lines are obtained with the nuclear potential which yields the barrier height of 
$B$=166 MeV and 179.9 MeV, respectively. }
\label{fig:CC_P1_P2_Ni} 
\end{figure} 
 
%\begin{table}
%\caption{ Parameters of the nuclear potential used for fig. \ref{fig:CC_P1_P2_Ni} for the  $^{60}$Ni+$^{116}$Sn reaction)}
%\begin{center}
%\begin{tabular}{c|c|c|c}
%\hline\hline
%&$V_0$ (MeV) & $r_0$ (fm) & $a_0$ (fm) \\
% \hline
%Realistic barrier & 116 & 1.2 & 0.687   \\
%Suppressed fusion &  80 & 1.1 & 0.487 \\
%\hline\hline
%\end{tabular}
%\end{center}
%\label{tab:Ni_Sn_param}
%\end{table}

%Because there is no experimental data for the capture cross section of the reaction $^{60}$Ni+$^{116}$Sn, 
%we can only study qualitatively the effect of the capture on the probability to transfer one and two neutrons. 

We next discuss the role of absorption in the transfer reactions by fully taking into account the 
 tunneling effect. For this purpose, we 
still use the three-channel problem in the previous subsection. 
Figure \ref{fig:CC_P1_P2_Ni} compares two calculations for the transfer probabilities for the 
$^{60}$Ni+$^{116}$Sn system. The first calculation shown by the dotted line is the same as 
the dashed line in Fig. \ref{fig:comp_SC_cc_ratio}. This result is obtained with the nuclear potential 
of Table V, which yields the barrier height of $B=$ 179.9 MeV. 
Notice that the Broglia-Winther potential \cite{Bro91} leads to a somewhat lower barrier height, that is, 
$B=$ 166 MeV. 
We therefore repeat the transfer calculations using the potential with 
$V_0$=116 MeV, $r_0$=1.2 fm, and $a_0$=0.687 fm, which yields the same barrier height as the Broglia-Winther 
potential. 
The result so obtained is denoted by the solid line in the figure. 
One can see that the one-neutron transfer cross sections do not differ much between the two calculations. 
In contrast, it is interesting to notice that 
the two-neutron transfer cross sections are considerably affected by the choice of 
the nuclear potential. 
Notice that we use the same imaginary potential for all the channels. 
Even so, we reach the same conclusion as in Fig. 
\ref{fig:sim_Q0_fus}, that is, the 2n transfer amplitude is more absorbed than the 1n transfer amplitude.
In this case, the absorption is more likely due to the capture rather than the inelastic processes. 

\begin{figure}[!ht] 
\centering\includegraphics[width=\linewidth]{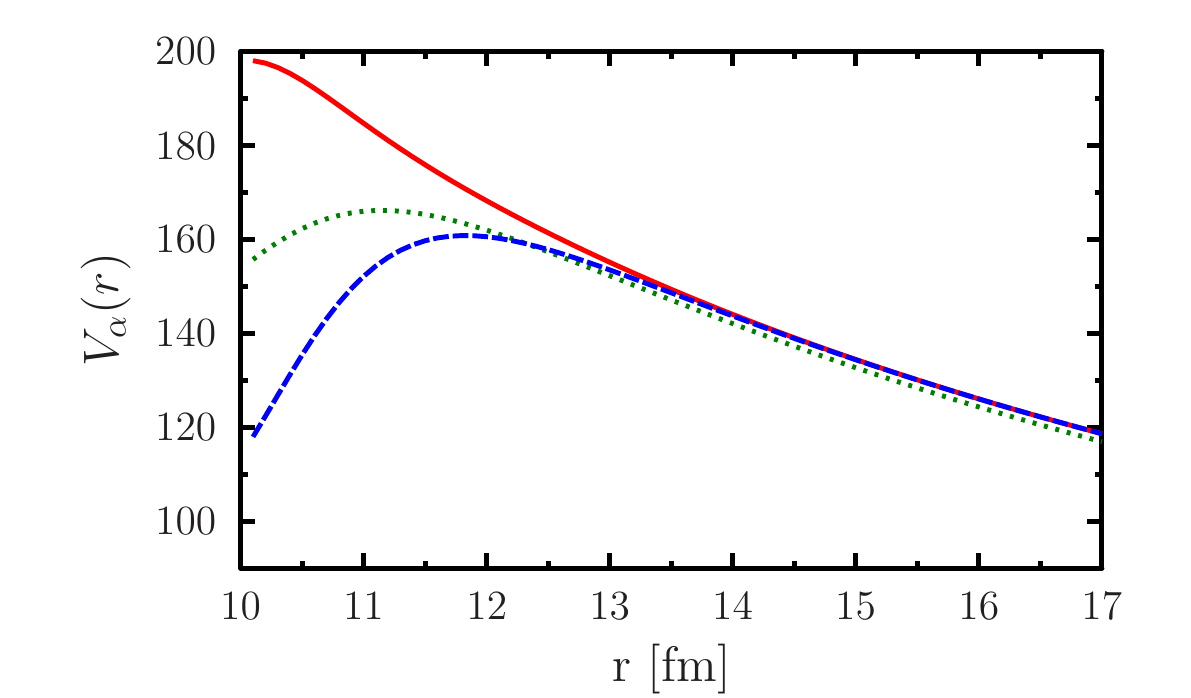}
\caption{ (Color online) The three eigen-barriers obtained by diagonalizing the intrinsic Hamiltonian 
matrix, Eq. \eqref{eq:Hamilt}, at each internuclear separation, $r$. }
\label{fig:barr} 
\end{figure}  
 
In order to understand a reasoning for this phenomenon, Fig. \ref{fig:barr} 
shows the eigen-barriers by diagonalizing the coupling matrix, Eq. \eqref{eq:Hamilt}, at 
each position $r$. 
That is, 
\begin{equation}
V_\alpha(r)=V_N(r)+\frac{Z_PZ_Te^2}{r}+\lambda_\alpha(r),
\end{equation}
where $\lambda_\alpha(r)$ is an eigen-value of 
the coupling matrix.  
In general, the eigen-vectors depend on the position $r$ \cite{HT12}. 
However, for the direct two-neutron transfer scheme with $V_{01}=V_{12}$ and $Q_2$=0, 
one of the eigen-vectors becomes independent of $r$. This special eigen-vector has a structure 
of $|\psi_1\rangle = \frac1{\sqrt2} ( | 0n \rangle - | 2n \rangle ) $ with the eigenvalue of $-V_{02}(r)$. 
Notice that the 1n channel does not contribute to this eigen-state. 
The other two eigen-states are given as a liner superposition 
of all the three channels, 
$| 0n \rangle$, $| 1n \rangle$, and $| 2n \rangle$. For a positive value of the ratio $\beta_{02}/a_{02}$, that is 
the case in our calculations, the eigenbarrier corresponding to the state 
$|\psi_1\rangle$ provides the lowest barrier among the three eigenbarriers, and is shown by the dashed line 
in Fig. \ref{fig:barr}. 
This implies that the capture occurs more easily from the 2n channel as compared to the 1n channel due to 
the lowest eigenbarrier, in which the 1n channel is absent. 

As we will discuss in the next section, the non-reproduction 
of the experimental probabilities in the vicinity of the barrier 
shown in Fig. \ref{fig:CC_P1_P2_Ni} may be attributed to the 
fact that we do not include the excited collective states in 
these calculations.
%It should be noted that the non reproduction of the experimental data in fig. \ref{fig:CC_P1_P2_Ni} is due to the non consideration of the excited collective states.

\subsection{Simultaneous description of fusion and 
multi-neutron transfer}
 
\begin{table}
\caption{The Coulomb deformation parameter, $\beta_C$, the nuclear deformation parameter, $\beta_N$, 
the multipolarity and the parity, $\lambda^\pi$, and 
the excitation energy $E$ for the collective states included in the coupled-channels 
calculations. The radius parameter of $r_0$=1.2 fm is used in the coupling potentials. }
\begin{center}
\begin{tabular}{c|c|c|c|c}
\hline\hline
Nucleus & $\lambda^\pi$ & $\beta_{C}$ & $\beta_N$ & $E$ (MeV) \\
 \hline
$^{40}$Ca & $3^-$ & 0.43 & 0.43 & 3.737   \\
$^{96}$Zr & $3^-$ & 0.27 & 0.305 & 1.89   \\
\hline\hline
\end{tabular}
\end{center}
\label{tab:Ca_Zr_exc}
\end{table}
 
One of the most important issues in the study of transfer reactions is to investigate whether the same 
strengths for the transfer couplings simultaneously account for fusion and transfer cross sections. 
Such attempt has been successfully made for 
the $^{33}$S+$^{90,91,92}$Zr systems \cite{Cor90,Cor93}. 
We do here a similar attempt for the
$^{40}$Ca+$^{96}$Zr reaction, for which fusion cross sections have been measured with 
high precision \cite{Ste14,Tim97}. 
To this end, we include both the multi-neutron transfer channels and the collective inelastic channels 
in the coupled-channels equations. 
To be more specific, 
we take into account the one octupole phonon excitation in 
$^{40}$Ca as well as the octupole phonon excitations 
in $^{96}$Zr up to the three-phonon states. We include all the possible mutual excitations. 
The parameters for the collective couplings are given in Table \ref{tab:Ca_Zr_exc}. 
The deformation parameters for the Coulomb couplings are estimated with the measured $B(E3)$ values, while 
we slightly increase the nuclear deformation parameter for $^{96}$Zr in order to 
better reproduce the experimental data. 
The multi-neutron transfer channels are taken into account up to three-neutron transfer, 
with the method of Esbensen and Landowne \cite{Esb89}. That is, we treat the collective excitations and the 
transfer channels as independent degrees of freedom so that the channel wave functions are specified as 
$|n_{\rm tr}n_{\rm inel}\rangle$, where $n_{\rm tr}$ (= 0, 1, 2, and 3)  
indicates the number of transferred neutron while 
$n_{\rm inel}$ specifies the inelastic channels. 
This lead to the total number of 32 channels (= 2 $\times$ 4 $\times$ 4, where the first 2 is for the 
excitation in $^{40}$Ca, the second 4 is for the excitations in $^{96}$Zr, 
%excitations in $^{40}$Ca, that is, 0$^+$ and 3$^-$, the second 4 is for the excitations in $^{96}$Zr, 
and the last 4 is for the transfer channels). 
For the one- and the two-neutron transfers,  
the parameters for the transfer couplings 
are readjusted to fit the measured transfer probabilities 
with the direct two-neutron transfer scheme with the optimum $Q$-values, that is, $Q$=0. 
The resultant coupling coefficients are 
$\beta_{01}=-71$ MeV\,fm, $a_{01}$=1.13 fm, $\beta_{02}=-105$ MeV\,fm, 
and $a_{02}$=0.82 fm. 
The three-neutron transfer channel is included with the coupling $V_{23}$=$V_{01}$, $V_{13}$=$V_{02}$ 
and $V_{03}$=0. We employ $Q_3$ = +4 MeV 
in order to reproduce the experimental data for the three-neutron transfer reaction. 
We use the nucleus-nucleus potential given in Table \ref{tab:comp_SC_cc_ratio}.
 
\begin{figure}[tb] 
\centering\includegraphics[width=\linewidth]{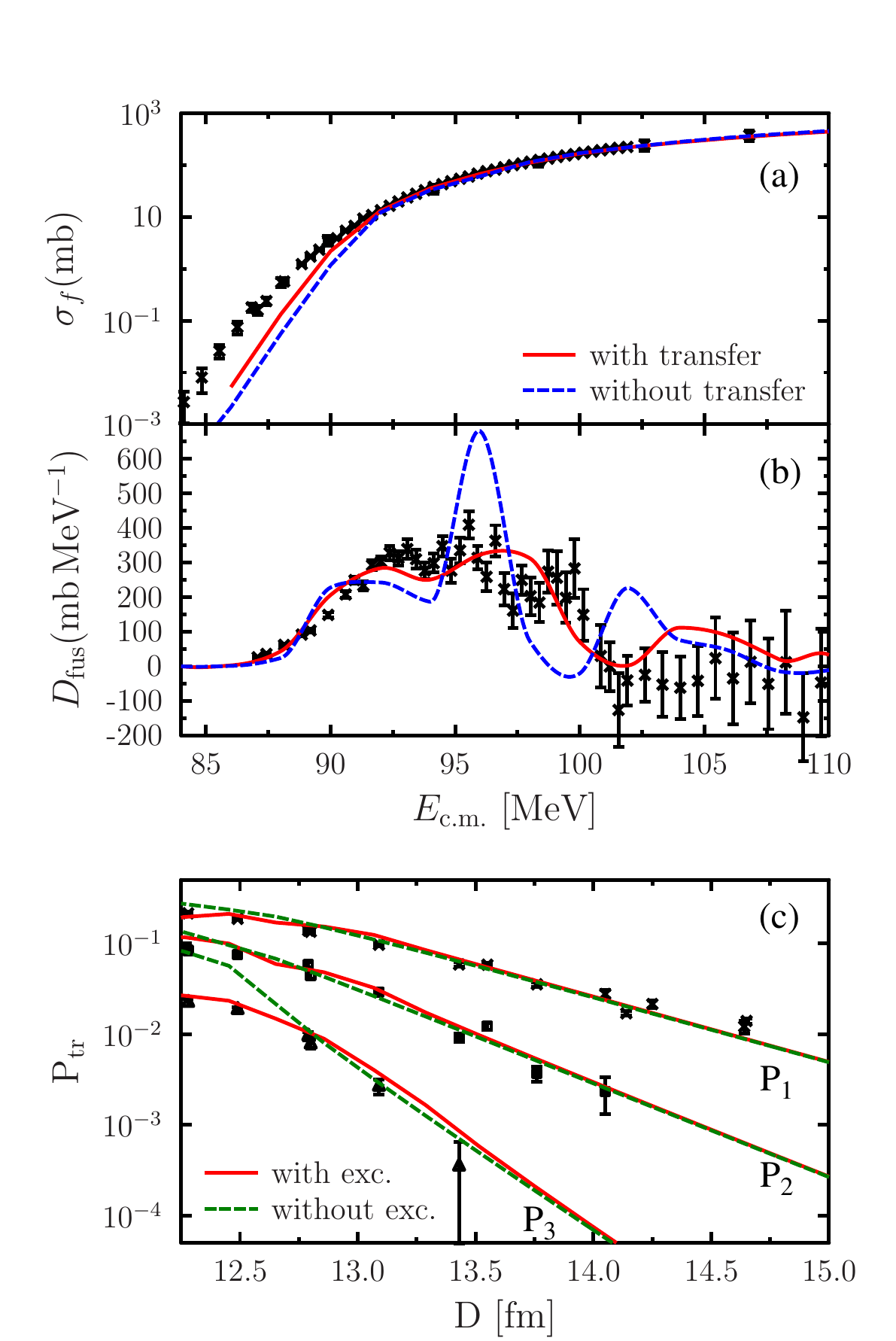}
\caption{(Color online)  (a) The fusion cross sections for the $^{40}$Ca+$^{96}$Zr reaction. 
The dashed line shows the results of the coupled-channels calculations including only the collective excitations in 
the colliding nuclei, while the solid line shows those with both the inelastic and the multi-neutron 
transfer channels. (b) The corresponding fusion barrier distribution. The experimental data are taken from 
Refs. \cite{Ste14,Tim97}. (c) The probabilities for the multi-neutron transfer reactions 
obtained with (the solid lines) and without (the 
dashed lines) taking into account the inelastic excitations. The experimental 
data are taken from Ref. \cite{Cor11}. }
\label{fig:fus_trsf} 
\end{figure}  

Figures \ref{fig:fus_trsf} (a) and \ref{fig:fus_trsf} (b) 
show the results of the coupled-channels calculations so 
obtained for the fusion cross sections $\sigma_{\rm f}$ 
and the fusion barrier distributions $D_{\rm fus}$ for the $^{40}$Ca+$^{96}$Zr 
system, respectively. 
Here, the fusion barrier distribution is defined as 
$D_{\rm fus}=d^2(E \sigma_{\rm f})/dE^2$ \cite{RSS91,DHRS98}. 
In the figures, the dashed lines denote the results obtained 
by including only the inelastic excitation in the 
colliding nuclei, while the solid lines are obtained by including in addition the multi-neutron transfer 
couplings. One can see that both the fusion cross sections and the barrier distributions are well 
reproduced by including the multi-neutron transfer channels, although fusion cross sections are still 
underestimated at low energies. 
In particular, the flatness of fusion barrier distribution is well 
reproduced by this calculation. 
We do not know the origin for the underestimate of fusion cross sections at low energies, but a similar 
tendency is seen also in another coupled-channels calculation reported in Ref. \cite{Esb14}. 
This could be due to the absence of other transfer channels, such as proton and alpha particle transfers, 
or the change in the collective couplings for the transfer channels. 

The transfer probabilities are shown in Fig. \ref{fig:fus_trsf} (c). 
The solid lines show the results of the full coupled-channels calculations, while the dashed lines are 
obtained by excluding the inelastic channels, that is, by 
including only the multi-neutron transfer channels. 
One can see that the inelastic excitations do not affect much 
the one- and two-neutron transfer channels, but 
the three-neutron transfer channel is sensitive to the inelastic excitations, especially 
for $D$$<$13 fm.   
One can also see that the present coupled-channels calculations well reproduce the experimental data for the 
transfer probabilities. 
It is worthwhile to mention that the fusion cross sections and the transfer probabilities are reproduced simultaneously within 
the single framework.

\section{Summary}
 
We have carried out a phenomenological study on the heavy-ion 
multi-neutron transfer reactions using the coupled-channels approaches. 
The aim was to adjust the parameters for the transfer couplings 
using the experimental data at energies far below the Coulomb 
barrier, where the perturbation treatment is applicable, 
and to investigate the dynamics at energies around the barrier, 
where the higher order terms as well as the absorption are important.
We have applied this strategy to the 
$^{40}$Ca+$^{96}$Zr and $^{60}$Ni+$^{116}$Sn systems, for which the 
transfer probabilities were recently measured. 
We have first used the semi-classical approximation to a three-channel 
problem with one- and two-neutron transfer channels,
and have obtained the following conclusions: 
i) the inclusion of the direct coupling between the entrance 
channel and the two-neutron transfer channel is necessary 
in order to reproduce the experimental data, ii) the higher order 
dynamics is important at energies around the Coulomb barrier and 
the first order treatment may not be sufficient, iii) 
the absorption has an important effect on the transfer 
probabilities at energies around the Coulomb barrier, and iv) 
the absorption plays a more important role for the two-neutron 
transfer channel as compared to the one-neutron transfer channel. 

The role of absorption has been confirmed also by 
using the quantal  
coupled-channels method for the $^{60}$Ni+$^{116}$Sn system. 
For the $^{40}$Ca+$^{96}$Zr system, 
we have succeeded to reproduce simultaneously 
the fusion cross sections, the fusion barrier distribution, 
and the transfer probabilities up to the three-neutron transfer 
channel, by including both the collective excitations and the 
transfer channels in the 
coupled-channels calculations. 

There have not been many systems for which both 
the fusion and the multi-neutron 
transfer cross sections are available at energies around and far 
below the Coulomb barrier. 
In that sense, 
the present phenomenological analysis is somewhat limited. 
It would be interesting to test the present approach to future 
experimental data for fusion and multi-neutron transfer reactions. 
 
\section*{Acknowledgement}
 
We thank 
D. Lacroix, 
D. Montanari, L. Corradi and A. Vitturi for useful discussions.
G.S. thanks also G. Adamian and N. Antonenko for useful discussions. 
G.S. acknowledges the Japan Society for the Promotion of Science
for the JSPS postdoctoral fellowship for foreign researchers.
This work was supported by Grant-in-Aid for JSPS Fellows No. 14F04769.

\end{document}